\documentclass[10pt,journal]{IEEEtran}
\usepackage{cite}
\usepackage{graphicx}
\usepackage{amsmath}
\usepackage{float}
\usepackage{xcolor}
\usepackage{comment}
\usepackage{dblfloatfix}    
\usepackage{pbalance}

\usepackage{fancyhdr}
\fancypagestyle{firstpage}
{
  \fancyhf{}
  \fancyfoot[C]{\footnotesize \copyright 2025 IEEE. All rights reserved, including rights for text and data mining, and training of artificial intelligence and similar technologies. Personal use is permitted, but republication/redistribution requires IEEE permission. See https://www.ieee.org/publications/rights/index.html for more information. \textit{DOI: 10.1109/TDMR.2025.3576042}}
}

\begin{document}

\title{Engineering Wake-Up-Free Ferroelectric Capacitors with Enhanced High-Temperature Reliability}
\author{Nashrah Afroze$^{1,\wedge}$, Salma Soliman$^1$, Yu-Hsin Kuo$^1$, Sanghyun Kang$^1$, Mengkun Tian$^2$, Priyankka Ravikumar$^1$, Andrea Padovani$^3$, Asif Khan$^{1,4,\$}$ \newline
$^1$School of Electrical and Computer Engineering, Georgia Institute of Technology, GA, USA; $^2$Institute of Materials and Systems, Georgia Institute of Technology, GA, USA; $^3$Department of Engineering Sciences and Methods (DISMI), University of Modena and Reggio Emilia, Reggio Emilia, Italy; $^4$School of Materials Science and Engineering, Georgia Institute of Technology, GA, USA. \\$\{${$^{\wedge}$nafroze3,$^{\$}$akhan40}$\}@$gatech.edu%

\thanks{\copyright 2026 IEEE. All rights reserved, including rights for text and data mining, and training of artificial intelligence and similar technologies. Personal use is permitted, but republication/redistribution requires IEEE permission. See https://www.ieee.org/publications/rights/index.html for more information. \textit{DOI 10.1109/TDMR.2026.3679333}}
}

{\let\newpage\relax\maketitle}


\begin{abstract}
We systematically explore the design space of ferroelectric hafnium–zirconium oxide (H\textsubscript{0.5}Z\textsubscript{0.5}O or HZO) heterostructures for reliable high-temperature operation. HZO films are deposited using thermal and plasma-enhanced atomic layer deposition (Th-ALD and PE-ALD) on tungsten (W) and titanium nitride (TiN) bottom electrodes (BE), while maintaining identical top electrodes. We demonstrate that PE-ALD HZO capacitors integrated with W BE exhibit wake-up-free switching up to 125~$^\circ$C, along with significantly improved endurance compared to Th-ALD HZO/W devices across a wide temperature range (85–125~$^\circ$C). By decoupling the contributions of the plasma-deposited HZO film and the oxidized bottom interface inherently formed during PE-ALD, we identify the oxidized W interfacial layer (WO\textsubscript{x}) as the primary factor governing endurance enhancement and wake-up suppression at elevated temperatures, while the PE-ALD HZO film provides secondary benefits in reducing wake-up. In contrast, PE-ALD HZO capacitors fabricated on TiN BE show no substantial improvement in wake-up behavior or endurance relative to Th-ALD HZO/TiN devices, despite the formation of an unintentional TiO\textsubscript{x}N\textsubscript{y} interfacial layer, and instead exhibit degraded polarization. This difference arises from the significantly weaker endurance enhancement and no wake-up suppression provided by oxidized TiN compared to oxidized W under comparable oxidation conditions. Overall, PE-ALD HZO films enable superior ferroelectric performance at elevated temperatures only when deposited on W BE, while Th-ALD HZO films remain a viable option for high-temperature operation on TiN BE. These findings clarify the interplay between deposition technique, electrode chemistry, and interfacial oxidation, and provide design guidelines for integrating ferroelectric memories into monolithic 3D systems under stringent thermal constraints.

\end{abstract}


\section{\textbf{Introduction}}


 \begin{figure}[!b]
    \centering
    \includegraphics[scale=0.6]{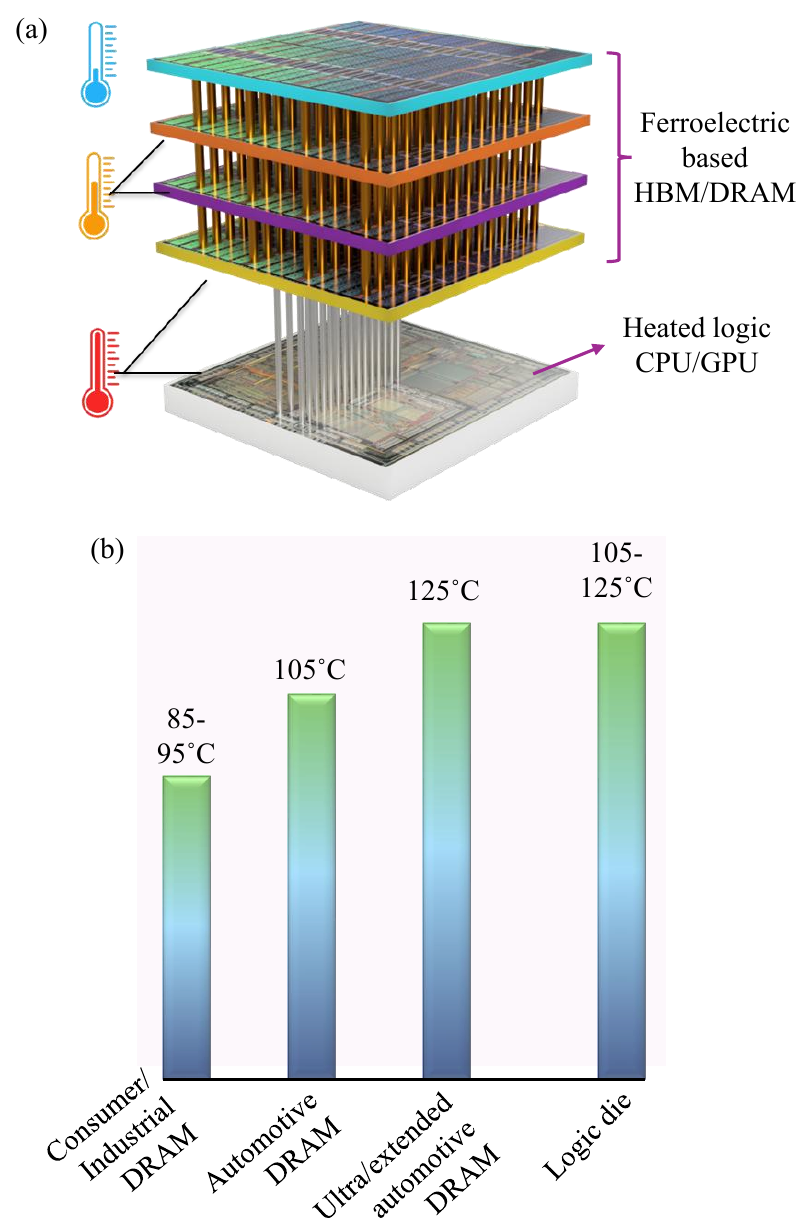}
    \caption{(a) Schematic representation of the temperature profile of an advanced 3-D stacked logic-on-memory system showing the importance of  reliability and stability of ferroelectric memory technology at higher temperature for 3D integration. (b) Operating temperatures of Micron
DRAMs based on different applications.}
    \label{fig:enter-label}
\end{figure}

Artificial Intelligence (AI) is accelerating advances in  high-performance computing, cloud infrastructure, healthcare and robotics, but data-heavy applications like autonomous driving and augmented reality are pushing memory systems to their limits. This highlights the need for high-performance, thermally reliable memory technologies. Ferroelectric memories have emerged as strong candidates for fast, dense, and ultra-low energy non-volatile  device for memory and storage technologies\cite{Asif_Nature, Muller_IEDM}, with recent demonstrations of ultra-dense 32 GB 3D-integrated FRAM\cite{Micron_1}. However, reliability still imposes serious challenges in their application\cite{Chavan_IMW}. In next-generation heterogeneous and monolithic 3D (H3D and M3D) systems, scaling of 3D integration subjects the inner dies to thermal stress due to increased stack height, heat crosstalk, and distance from heat sinks\cite{Zhang_3DIC} (Fig. 1a). In high-performance computing systems, logic dies can reach elevated junction temperatures in the range of approximately 105–125\textdegree C under sustained or peak workloads \cite{Zhou_3D}. To evaluate long-term reliability under such thermal conditions, memory devices are typically subjected to accelerated high-temperature operating life (HTOL) testing at 125\textdegree C under electrical bias, in accordance with the JEDEC JESD22-A108 standard.


\begin{figure}[!htb]
    \centering
    \includegraphics[scale=0.34]{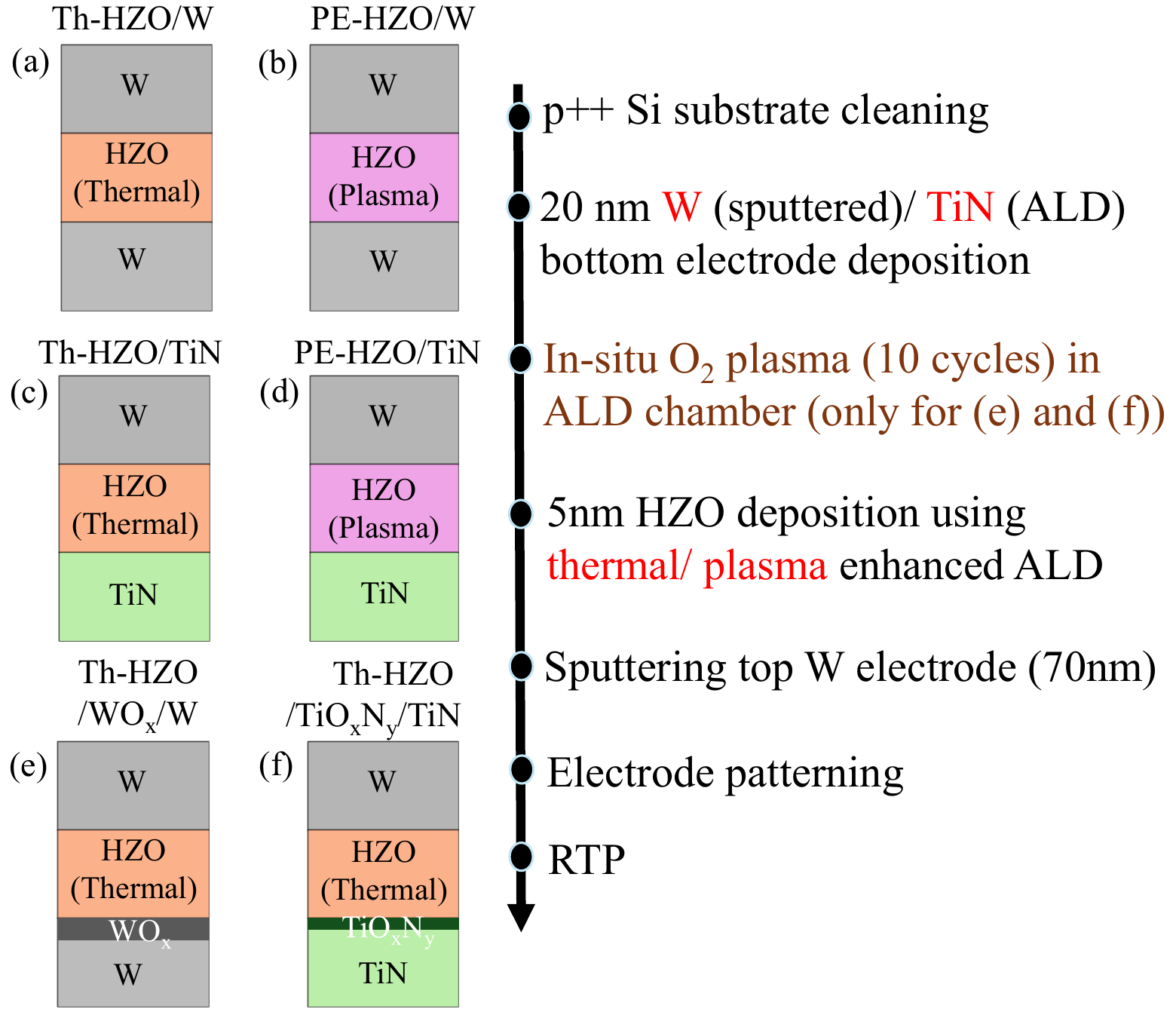}
    \caption{Device structures of ferroelectric capacitors with W (a-b) and TiN (c-d) BE along with the device fabrication process flow. Bottom TiN electrode is deposited using ALD followed by HZO deposition without breaking the vacuum. (e-f) 10 cycles of O\textsubscript{2} plasma are used to oxidize the bottom electrode interface followed by Th-ALD deposited HZO films.   }
    \label{fig:enter-label}
\end{figure}

\begin{figure}[!htb]
    \centering
    \includegraphics[scale=0.6]{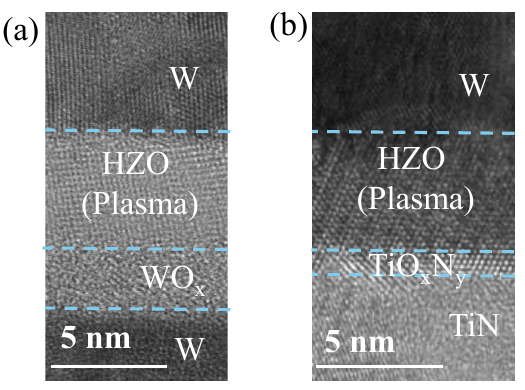}
    \caption{STEM image of PE-HZO devices with W (a) and TiN (b) bottom electrodes. Oxidized bottom electrode interfaces creating WO\textsubscript{x} (a) and TiO\textsubscript{x}N\textsubscript{y} (b) are clearly visible.  }
    \label{fig:enter-label}
\end{figure}

\begin{figure*}[!htb]
    \centering
    \includegraphics[scale=0.5]{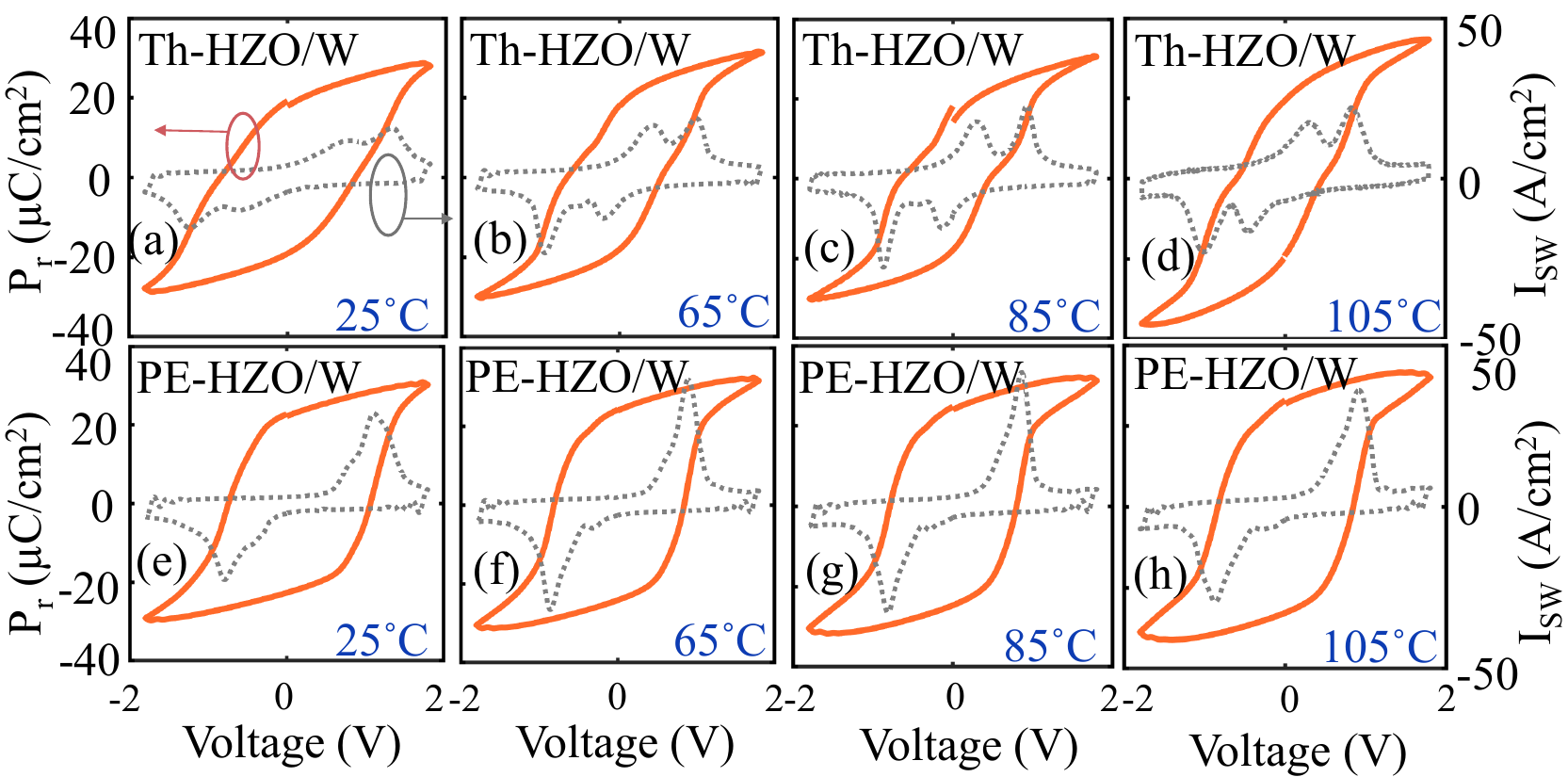}
    \caption{ P-V and I\textsubscript{sw}-V of Th-HZO/W (a-d) and PE-HZO/W (e-h) devices at pristine state measured at 25\textdegree, 65\textdegree, 85\textdegree and 105\textdegree C respectively.    }
    \label{fig:enter-label}
\end{figure*}

Figure 1(b) presents the maximum operating temperature specifications of Micron DRAM products across various application domains, illustrating the increasing thermal demands placed on memory technologies\cite{Micron}. For comparison, the maximum operating temperature of non-volatile memories such as NAND flash is also typically around 105\textdegree C\cite{NAND}. These specifications underscore the necessity for robust high-temperature memory operation to enable reliable integration of ferroelectric memories in emerging heterogeneous and monolithic 3D (H3D and M3D) platforms, as well as memory-on-logic architectures. However, Ferroelectric thin films typically suffer from degradation in ferroelectricity, enhanced wake-up effects and poor endurance at elevated temperatures.\cite{Mittmann_IEDM, Shroeder_Adv, Sunbul_Adv, Park_Adv, Mimura_APL}.

 Plasma-enhanced atomic layer deposition (PE-ALD) of hafnium–zirconium oxide (HZO) has attracted significant interest due to its high remanent polarization, wake-up-free behavior at room temperature, compatibility with low thermal-budget BEOL processes, and moderate endurance compared to thermally deposited ALD (Th-ALD) HZO films~\cite{Onaya_ME2019,Hur_Nanotech,Park_JEDS,Hsain,Chen_ACSAMI}. Despite these advantages, a comprehensive comparison of the performance and reliability of PE-ALD and Th-ALD HZO films under elevated operating temperatures, particularly for memory-on-logic applications, remains largely unexplored.

Although numerous studies have reported improvements in polarization, wake-up behavior, and endurance at room temperature through electrode engineering~\cite{Chen_IEDM,Alcala_AFM,Wang_ACSAMI,Wang_Adv}, controlled oxygen supply during deposition~\cite{Kashir_ACSAEM}, and interface engineering~\cite{Yang_MSSP,Lee_AMI,Shi}, a systematic investigation of the combined effects of ALD deposition technique, bottom electrode material, and interfacial layer formation has been largely overlooked for both room and high-temperature operation.

In this work, we systematically compare the performance of hafnium–zirconium oxide (HZO) films deposited by thermal and plasma-enhanced atomic layer deposition (Th-ALD and PE-ALD) on tungsten (W) and titanium nitride (TiN) bottom electrodes for high-temperature memory-on-logic applications. We show that plasma-deposited HZO films (PE-HZO) exhibit superior performance at elevated temperatures compared to their thermally deposited counterparts (Th-HZO) when integrated with W bottom electrodes. In contrast, PE-HZO films do not provide substantial performance improvement over Th-HZO when deposited on TiN bottom electrodes. By decoupling the respective contributions of the plasma-deposited HZO film and the oxidized bottom interfacial layer, we identify the oxidized interface as a key factor governing the behavior of plasma-deposited capacitors, particularly at elevated temperatures. These findings provide critical insights into the interplay between deposition technique, electrode chemistry, and high-temperature reliability, and offer design guidelines for realizing robust ferroelectric memories for high-temperature memory-on-logic applications.

\section{\textbf{Experimental Details}}
W(TE)/HZO/W(BE) and W(TE)/HZO/TiN(BE) ferroelectric capacitors were fabricated on p\textsuperscript{+} Si substrates following the standard process flow shown in Fig. 2. 5 nm HZO was deposited either by plasma-enhanced ALD (PE-ALD) using O\textsubscript{2} plasma or by thermal ALD (Th-ALD) using H\textsubscript{2}O as the oxidant. Keeping the top electrode same (W), either sputtered W or ALD deposited TiN is used as bottom electrode. All the samples underwent post-deposition annealing at 550\textdegree C for 1 minute. Since the PE-ALD deposition process leads to partial oxidation of the bottom electrode, as evidenced in Fig.~3, it is essential to decouple the respective contributions of the PE-ALD HZO film and the oxidized bottom interfacial layer to the overall performance of the PE-HZO devices. To achieve this, a controlled oxidation of the bottom electrode was intentionally introduced by applying 10 cycles of O\textsubscript{2} plasma in the ALD chamber in the absence of Hf/Zr precursors, followed by the deposition of a 5~nm HZO film using Th-ALD (Fig. 2e-f). The remainder of the fabrication process followed the same flow as illustrated in Fig.~2. Comparison of the oxidized BE devices with Th-HZO devices isolates the effect of the oxidized bottom layer, while comparison with PE-HZO devices elucidates the contribution of the PE-HZO film.

The endurance of the capacitors was evaluated under bipolar cycling, wherein the device polarization was alternated between +$P_{sw}$ and -$P_{sw}$ over a voltage range of ±1.6 V to ±2.0 V and a temperature range of 25 $^\circ$C to 125 $^\circ$C (Fig. 7a). Cycling was carried out using 200 ns pulses at a frequency of 2.5 MHz, with periodic interruptions to perform P–V measurements using 20 $\mu$s triangular pulses. These measurements enabled extraction of polarization and switching current at different stages of stress cycling. All experiments were conducted using a Keysight B1500 semiconductor parameter analyzer. The corresponding results and analyses are presented in the following sections.

\section{\textbf{Results and discussions}}
\subsection{{High temperature wake-up with W BE}}

\begin{figure}[!htb]
    \centering
    \includegraphics[scale=0.38]{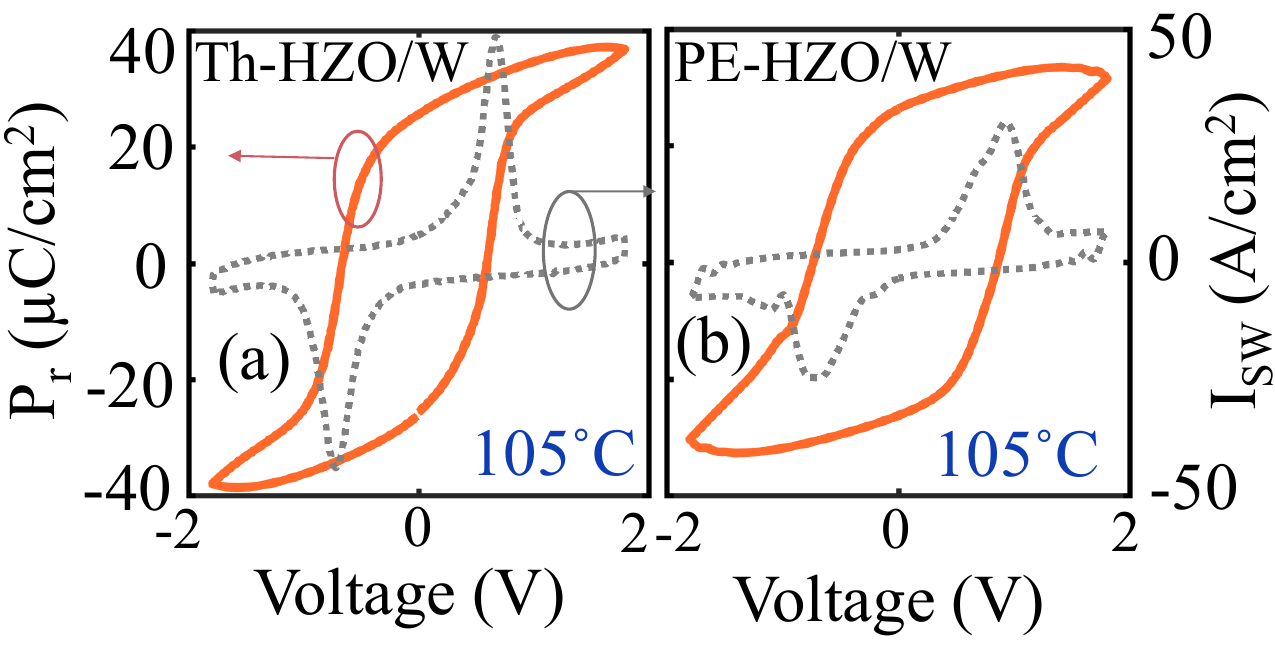}
    \caption{P-V and Isw-V of Th-HZO/W (a) and PE-HZO/W (b) at 105\textdegree C after 10\textsuperscript{5} endurance cycles (±1.8V/200ns). No significant leakage traits are observed.  }
    \label{fig:enter-label}
\end{figure}

Figure 4 shows the P–V and I\textsubscript{SW}–V characteristics of Th-ALD and PE-ALD deposited HZO capacitors on W BE (Th-HZO/W and PE-HZO/W devices respectively) in their pristine states, measured over a temperature range of 25-105\textdegree C. Both the devices enter major hysteresis loop at ±1.8V. However, polarization is higher in the PE-HZO/W device compared to Th-HZO/W as seen in figs. 4a,e due to having higher orthorhombic phase as confirmed by the deconvolution of the dominant o(111)/t(101) peak from the grazing incident X-ray diffraction (GI-XRD) scans (not shown). Th-HZO/W device shows typical ferroelectric behavior at room temperature (Fig. 4a). However, the switching behavior progressively shifts toward an antiferroelectric-like character, manifested by the emergence of double-peaked switching currents and increasingly pinched hysteresis loops as temperature increases up to 105\textdegree C (Fig. 4b-d). By contrast, the PE-HZO/W HZO device demonstrates remarkable thermal robustness: even at elevated temperatures, it consistently maintains a single-peaked switching current profile and a stable, purely ferroelectric hysteresis response (Fig. 4e–h). These findings clearly indicate that PE-HZO/W capacitors retain their pristine polarization and ferroelectric properties more effectively under thermal stress, underscoring their suitability for memory-on-logic integration where high-temperature reliability is essential.

As shown in Fig. 5a, the double-peaked switching current observed in the Th-HZO/W device at elevated temperature disappears after $10^{5}$ bipolar fatigue cycles, indicating a wake-up effect. Both the Th-HZO/W and PE-HZO/W devices show minimal leakage after 10\textsuperscript{5} cycles. Therefore, the observed increase in polarization in the P–V loops after $10^{5}$ cycles can be attributed to the enhancement of polarization associated with the wake-up process. 

\begin{figure}[H]
    \centering
    \includegraphics[scale=0.4]{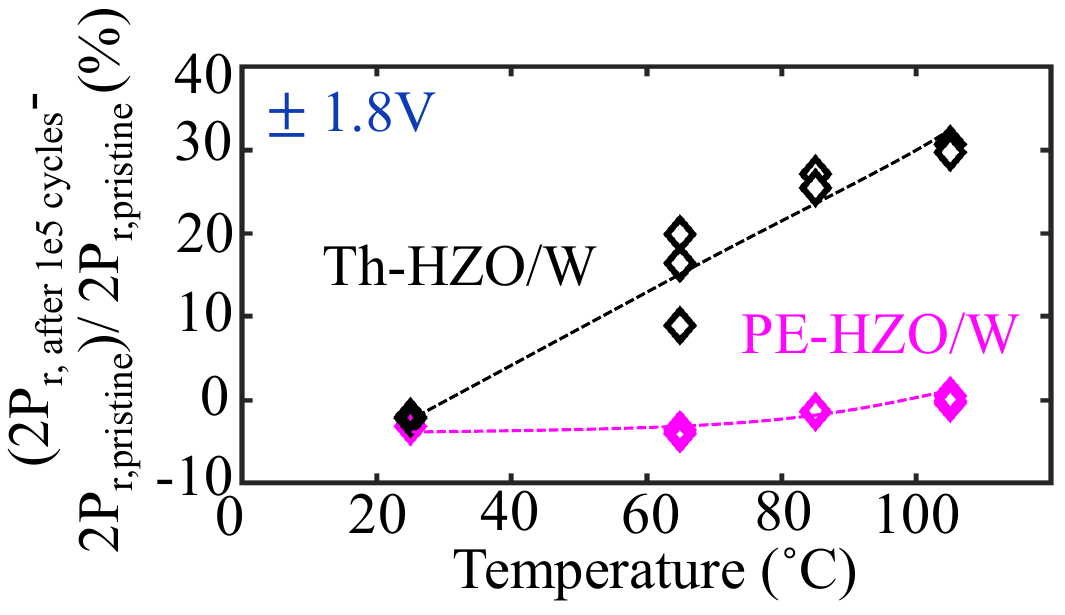}
    \caption{Increase in polarization after 10\textsuperscript{5} cycles compared to pristine state at different temperatures.}
    \label{fig:enter-label}
\end{figure}

\begin{figure*}[!htb]
    \centering
    \includegraphics[scale=0.55]{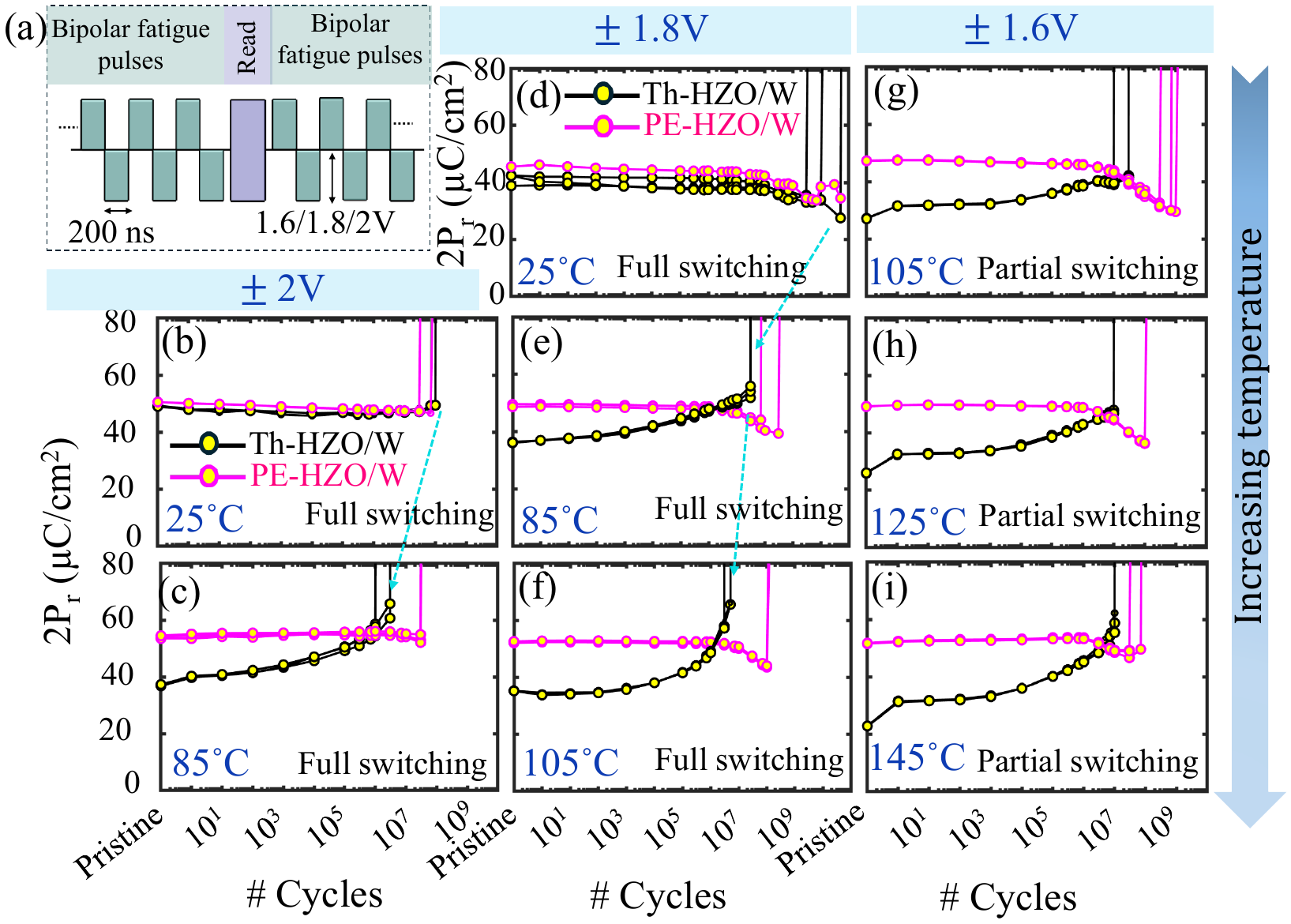}
    \caption{(a) Write endurance scheme. Note that applying ±1.8V leads to complete polarization switching (Fig. 4). (b-i) 2P$_r$ vs. cycles characteristics at ±2 V (first column, b-c), ±1.8 V (second column, d-f), and ±1.6 V (third column, g-i) at different temperatures.} 
    \label{fig:enter-label}
\end{figure*}

Figure 6 presents the relative increase in polarization after wake-up compared to the pristine state for both Th-HZO/W and PE-HZO/W devices at different temperatures, quantified as:   

\[
\frac{2P_{r,\text{after }10^{5}\,\text{cycles}} - 2P_{r,\text{pristine}}}{2P_{r,\text{pristine}}} \times 100
\]
For each condition, measurements were conducted on at least three devices to ensure statistical reliability. The results show a pronounced polarization increase in the Th-HZO/W devices after $10^{5}$ bipolar fatigue cycles (±1.8 V, 200 ns), whereas the PE-HZO/W  devices exhibit only a negligible change across the entire temperature range. This clear contrast highlights the strong suppression of the wake-up effect in PE-HZO/W device.

Since the PE-ALD deposition process leads to partial oxidation of the bottom electrode, as evidenced in Fig.~3, we further decoupled the role of oxidized BE (WO\textsubscript{x}) and the PE-HZO film on achieving the wakeup-free behavior by comparing Th-HZO/WO\textsubscript{x}/W device of Fig. 3e with Th-HZO/W device to understand the affect of oxidized interface, and with PE-HZO/W device to unveil the contribution of PE-HZO film itself. WO\textsubscript{x} inserted Th-HZO/W devices require approximately $10^{4}$ cycles to eliminate the double peaks in $I_{\mathrm{SW}}-V$ curve when measured under the same scheme as the other two device structures\cite{Afroze_ACS}, and exhibit polarization values in the pristine state comparable to those of Th-HZO/W devices, as shown in prior report~\cite{Afroze_IEDM}. The wakeup cycles are further reduced to 0 upon using plasma deposited HZO film. These results indicate that the enhanced polarization and wake-up-free behavior at elevated temperatures observed in PE-HZO/W devices originate from the PE-ALD-deposited HZO film itself, facilitated by the oxidized bottom electrode.

\subsection{{High temperature endurance with W BE}}
Endurance measurements were performed as described in Section II, using the pulse scheme shown in Fig. 7a. Multiple devices of each type were characterized to account for device-to-device variability. At room temperature, both PE-HZO/W and Th-HZO/W capacitors exhibit comparable endurance and polarization under full switching conditions (±2 V and ±1.8 V cycling and read voltages) (Fig. 7b,d). With increasing temperature, however, the Th-HZO/W devices display a clear degradation in both endurance and polarization, as indicated by the cyan arrows in Figs. 7b–f. In contrast, PE-HZO/W devices consistently demonstrate superior endurance across all elevated temperature measurements. A similar trend is observed under partial switching conditions, where endurance was tested with ±1.6 V cycling at elevated temperatures (Fig. 7g–i). Notably, no evidence of a wake-up effect is observed in PE-HZO/W devices under any condition. These results collectively highlight the excellent thermal stability and reliability of PE-HZO/W capacitors compared to their Th-HZO/W counterparts.

 \begin{figure}[!htb]
    \centering
    \includegraphics[scale=0.45]{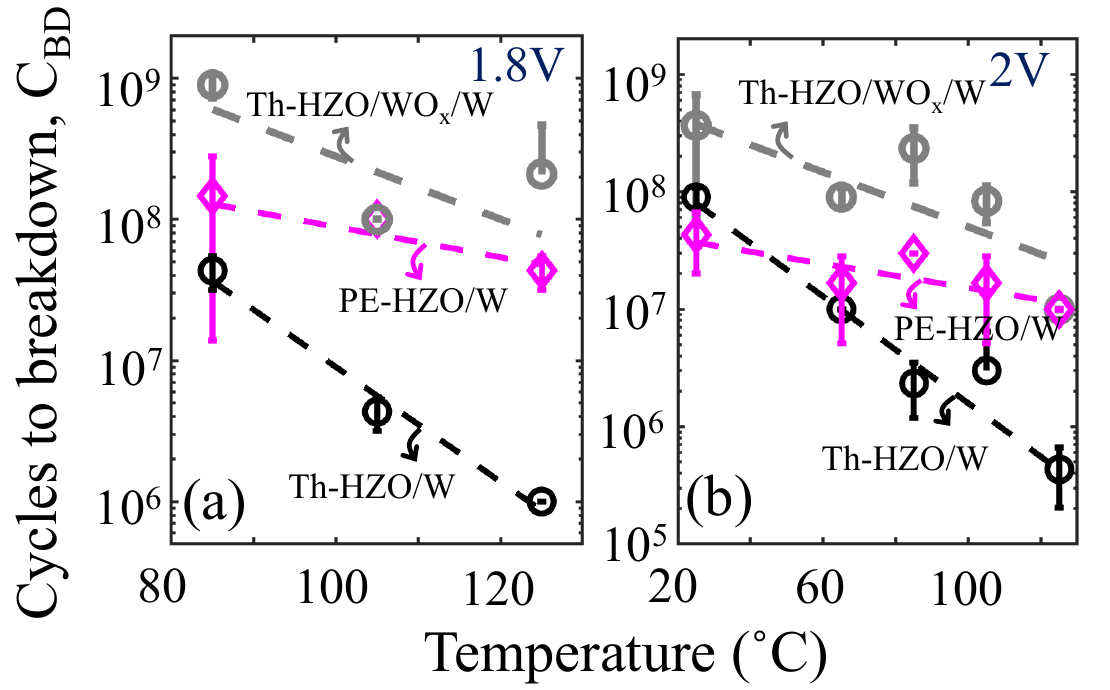}
    \caption{ Average cycles to breakdown (C$_{BD}$) vs temperature for Th-HZO/W (black), PE-HZO/TiN (magenta) and WO\textsubscript{x} inserted Th-HZO/W (gray) devices for 1.8V (a) and 2V (b) bipolar cycling. The dashed lines represent least-squares linear fits to the log\textsubscript{10}-transformed data.} 
    \label{fig:enter-label}
\end{figure}

Figure~8 summarizes the endurance characteristics of Th-HZO/W and PE-HZO/W HZO capacitors under full switching conditions across a wide range of operating temperatures. For Th-HZO/W devices, the cycles to breakdown (C\textsubscript{BD}) decrease sharply with increasing temperature, indicating a pronounced thermal sensitivity of endurance. In contrast, although PE-HZO/W devices exhibit slightly lower endurance than Th-HZO/W at room temperature (Fig. 8b), they retain their endurance much more effectively at elevated temperatures up to 125~$^\circ$C, highlighting their improved robustness under thermal stress and their relevance for high-temperature 3D integration applications. 

To decouple the contribution of the PE-ALD HZO film from that of the WO\textsubscript{x} interfacial layer, endurance measurements over a wide temperature range were performed on WO\textsubscript{x} inserted Th-HZO/W devices using the same measurement scheme, as indicated by the gray plots in Fig.~8. Notably, these devices exhibit significantly enhanced endurance under both 1.8~V and 2~V cycling compared to Th-HZO/W devices. However, the endurance of PE-HZO/W devices which naturally grew WO\textsubscript{x} remains lower than that of WO\textsubscript{x} inserted Th-HZO/W devices. These results indicate that the PE-HZO film itself does not contribute to improving high-temperature endurance and instead slightly degrades it compared to the Th-HZO film due to plasma-induced damage mechanisms \cite{PEALD}, consistent with previous reports \cite{Hur_Nanotech, Park_JEDS}. Rather, the endurance enhancement observed in PE-HZO/W devices compared to Th-HZO/W at elevated temperatures primarily arises from the WO\textsubscript{x} bottom interfacial layer, which enables a temperature-activated self-healing mechanism in ferroelectric capacitors, thereby extending device lifetime\cite{Afroze_IEDM, Afroze_IRPS}. The performance comparison of Th-ALD and PE-ALD HZO devices with W BE are presented in Fig. 12a.

 \begin{figure*}[!htb]
    \centering
    \includegraphics[scale=0.5]{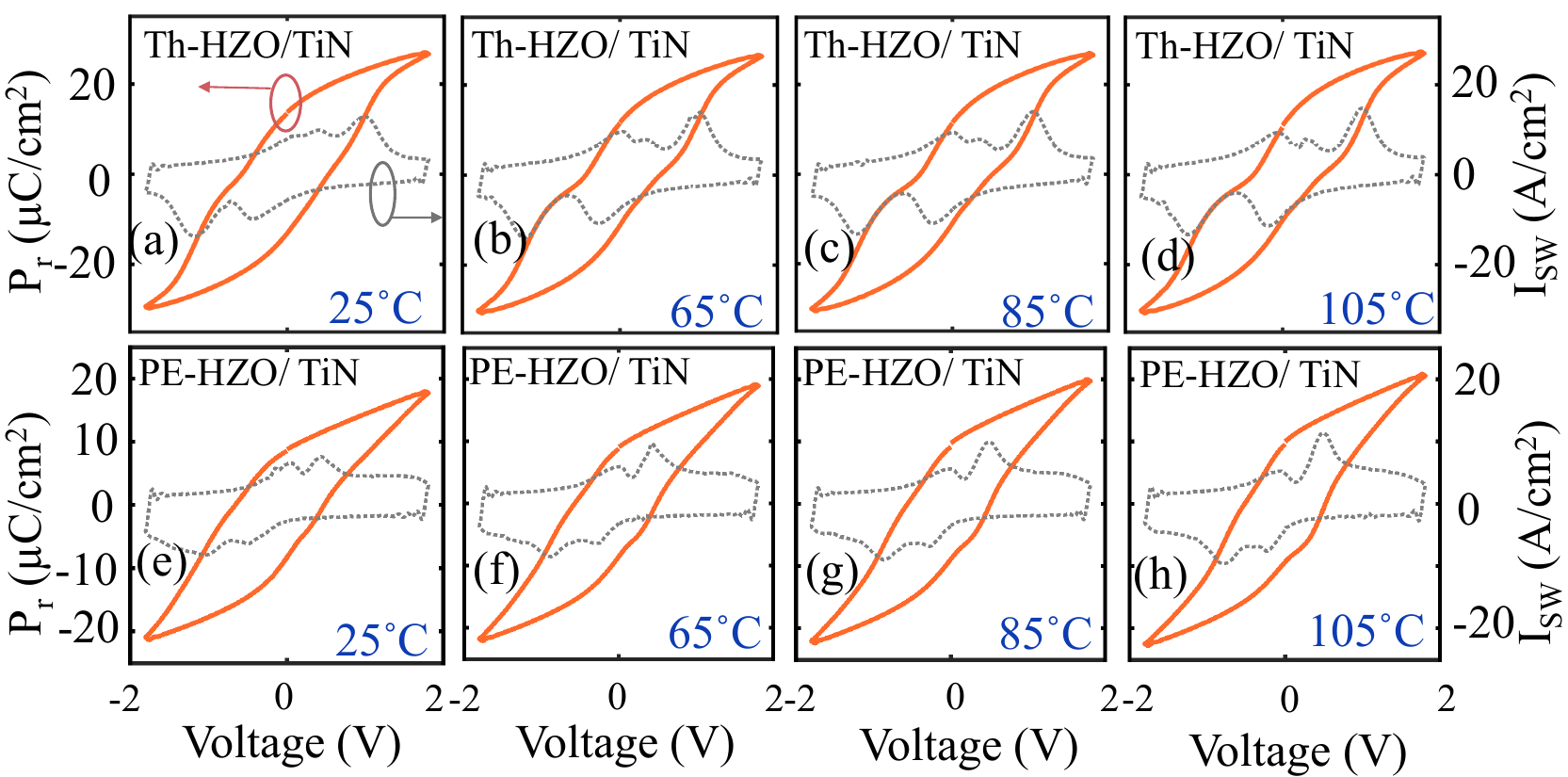}
    \caption{ P-V and I\textsubscript{sw}-V of Th-HZO/TiN (a-d) and PE-HZO/TiN (e-h) devices at pristine state measured at 25\textdegree, 65\textdegree, 85\textdegree and 105\textdegree C respectively.    }
    \label{fig:enter-label}
\end{figure*}

\subsection{{High temperature wake-up with TiN BE}}
To assess the generality of the impact of PE-HZO films on polarization and wake-up behavior, the bottom electrode in both Th-HZO/W and PE-HZO/W devices was replaced with TiN while keeping the TE and other fabrication steps unchanged, as illustrated in Fig.~2(c-d). Under this configuration, the PE-HZO/TiN structure does not exhibit wake-up-free behavior and instead shows pronounced double switching peaks in the pristine state at both room and elevated temperatures (Fig. 9e-h), similar to its thermal counterpart (Th-HZO/TiN), as shown in Fig.~9(a-d). Both PE-HZO/TiN and Th-HZO/TiN devices require approximately $10^{6}$ switching cycles to complete wake-up and suppress the double peaks in $I_{\mathrm{SW}}-V$, with the PE-HZO/TiN device exhibiting reduced polarization compared to the Th-HZO/TiN device (Fig.~10). The reduction in 2P\textsubscript{r} observed in the PE-HZO/TiN device is attributed to the adverse impact of the inherited oxidized interfacial layer introduced by the PE-ALD process, which inhibits templating of the o-phase of HZO on TiN and thereby hinders o-phase formation~\cite{Hsain}. These findings indicate that neither the presence of an oxidized TiN interfacial layer nor the PE-HZO film deposited on a TiN BE is sufficient to enable wake-up-free operation at elevated temperatures.

\subsection{{High temperature endurance with TiN BE}}
The endurance characteristics of TiN BE devices were evaluated across multiple devices using the same measurement scheme employed for the W BE devices, and the results are summarized in Fig.~11. Under 1.8~V bipolar cycling, PE-HZO/TiN devices do not exhibit any noticeable endurance enhancement at elevated temperatures compared to Th-HZO/TiN devices (Fig.~11a). In contrast, for 2~V bipolar cycling, although PE-HZO/TiN devices show poorer endurance than Th-HZO/TiN devices at room temperature, they begin to exhibit comparatively improved endurance at elevated temperatures, similar to PE-HZO/W devices (Fig.~11b).

To decouple the contributions of the oxidized bottom interface in PE-HZO/TiN devices from those of the PE-ALD HZO film itself, we have used Th-HZO devices with oxidized TiN BE (Th-HZO/TiO\textsubscript{x}N\textsubscript{y}/TiN) for comparison. The endurance behavior of these devices is also included in Fig.~11, indicated by the gray plots.

Compared to Th-HZO/TiN devices, Th-HZO/TiO\textsubscript{x}N\textsubscript{y}/TiN devices show enhanced endurance across all temperatures and cycling voltages, highlighting the TiO\textsubscript{x}N\textsubscript{y} layer as a potential oxygen reservoir over a wide temperature range\cite{Hsain}. However, the magnitude of this endurance improvement compared to Th-HZO/TiN is only $\sim 10\times$ at 125\textdegree C with 1.8V, smaller than that observed for Th-HZO/WO\textsubscript{x}/W devices relative to their Th-HZO/W counterparts ($\sim 10^{3}\times$). Owing to the relatively weaker contribution of the oxidized TiN interface to endurance enhancement, no significant improvement in endurance is observed in PE-HZO/TiN devices. These observations are consistent with the trends identified in W BE devices, where the oxidized bottom interface plays a key role in enhancing endurance at elevated temperatures, while the PE-HZO film reduces the extent of this enhancement when compared to Th-HZO devices with an oxidized BE. The performance comparison of Th-ALD and PE-ALD HZO devices with TiN BE are presented in Fig. 12b.

\begin{figure}[!htb]
    \centering
    \includegraphics[scale=0.38]{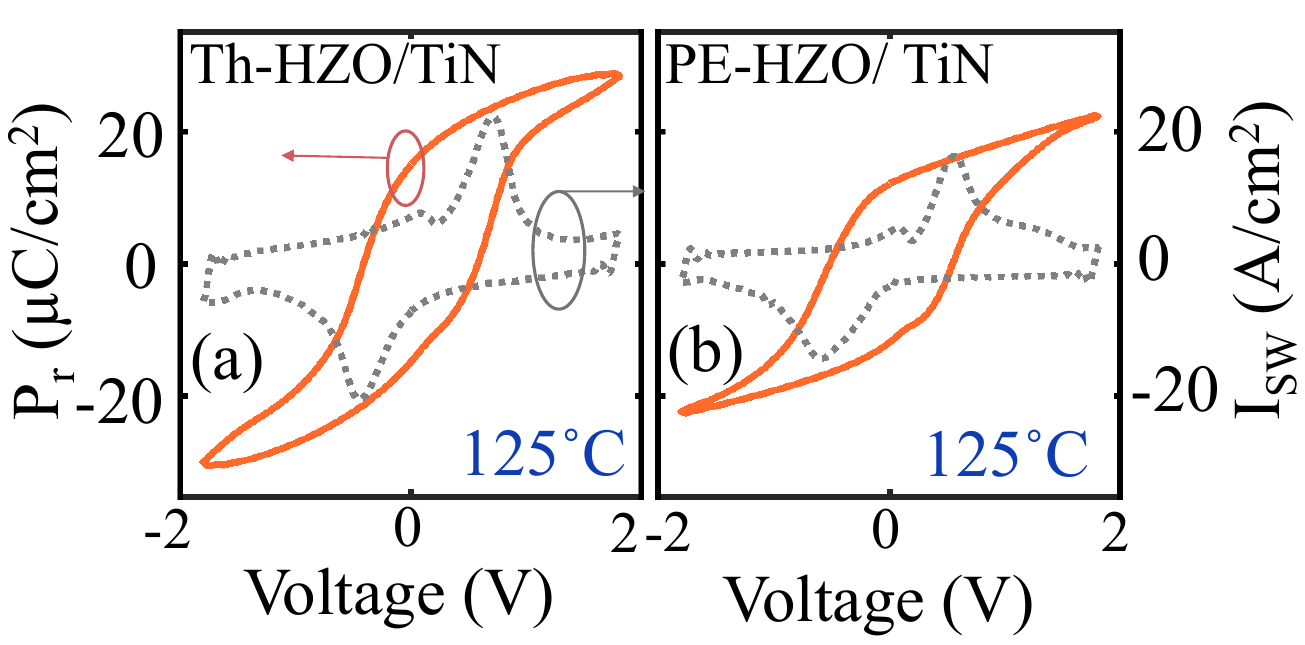}
    \caption{P-V and Isw-V of Th-HZO/TiN (a) and PE-HZO/TiN (b) at 125\textdegree C after 10\textsuperscript{6} endurance cycles (±1.8V/200ns). }
    \label{fig:enter-label}
\end{figure}

 \begin{figure}[!htb]
    \centering
    \includegraphics[scale=0.45]{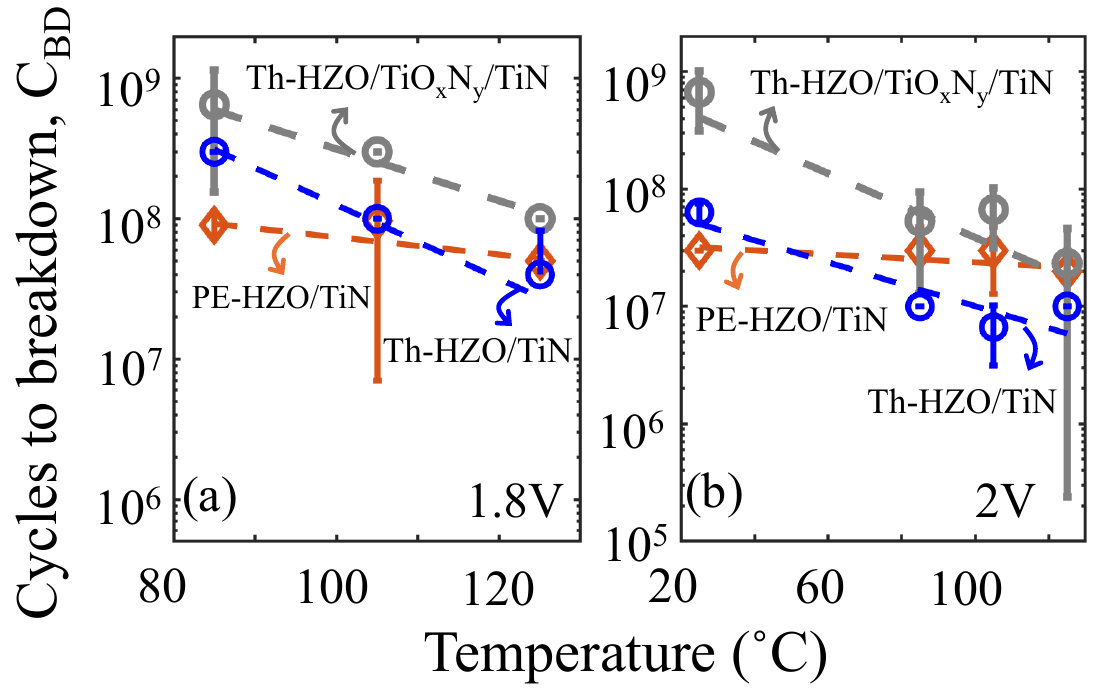}
    \caption{Average cycles to breakdown (C$_{BD}$) vs temperature for Th-HZO/TiN (blue), PE-HZO/TiN (brown) and Th-HZO/TiN with oxidized BE (gray) devices for 1.8V (a) and 2V (b) bipolar cycling. The dashed lines represent least-squares linear fits to the log\textsubscript{10}-transformed data.}
    \label{fig:enter-label}
\end{figure}

 \begin{figure}[!htb]
    \centering
    \includegraphics[scale=0.52]{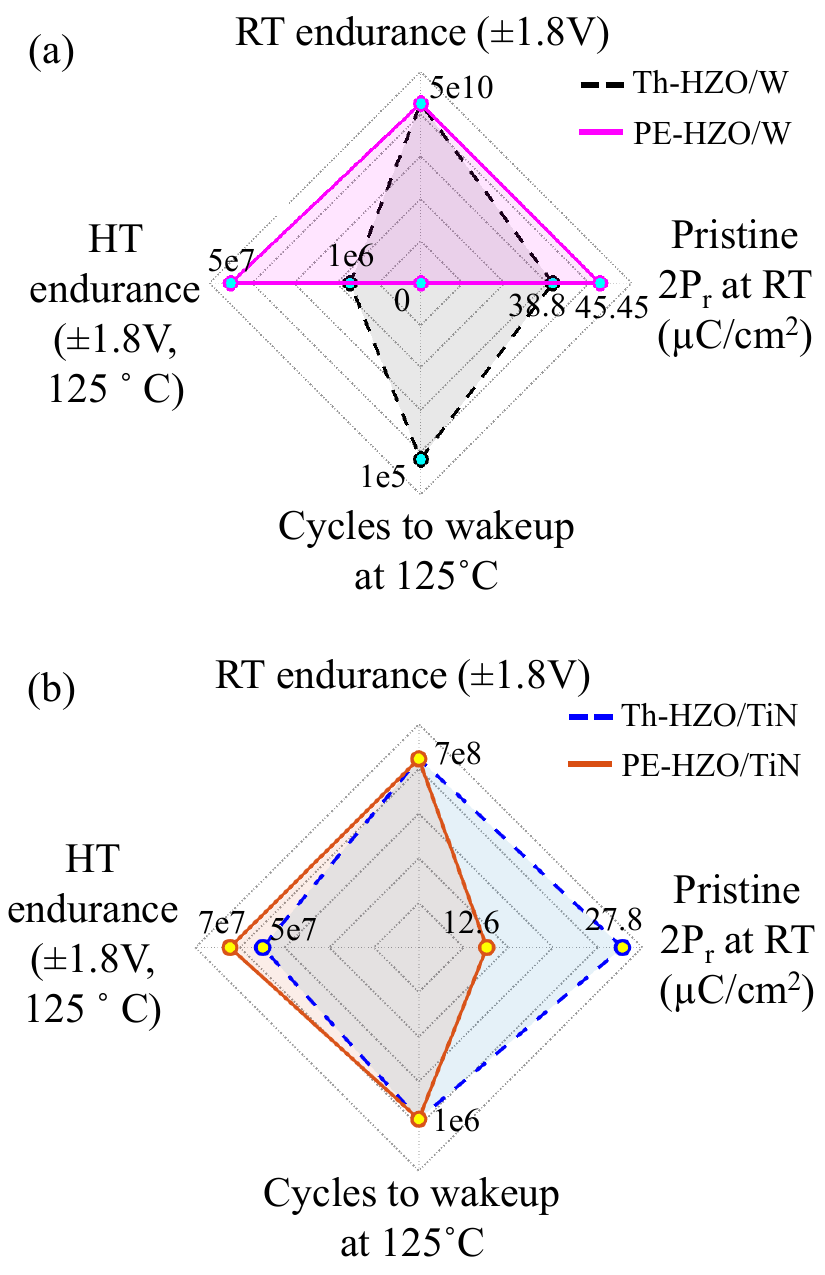}
    \caption{Spider plot showing comparison of 2P\textsubscript{r} at room temperature (RT), room temperature (RT) endurance for 1.8V bipolar cycling, endurance at 125\textdegree C for 1.8V bipolar cycling and cycles needed for wakeup at 125\textdegree C between Th-HZO and PE-HZO devices with W BE (a) and TiN BE (b).}
    \label{fig:enter-label}
\end{figure}

\section{\textbf{Conclusion}}

In conclusion, this work systematically elucidates the coupled roles of ALD deposition technique and bottom electrode chemistry in determining the high-temperature reliability of ultrathin HZO ferroelectric capacitors. By directly comparing Th-ALD and PE-ALD HZO films integrated with W and TiN bottom electrodes, we demonstrate that wake-up-free switching and enhanced endurance up to 125 °C are achieved exclusively in PE-ALD HZO devices employing W bottom electrodes. The endurance enhancement observed at elevated temperatures compared to Th-HZO/W counterpart primarily originates from the oxidized bottom interface formed during the PE-ALD HZO deposition process. In contrast, the PE-ALD HZO film itself contributes to suppressed wake-up behavior and enhanced polarization magnitude only when deposited on a W BE. This underscores the dominant role of the bottom electrode in governing ferroelectric behavior over a wide operating temperature range. The benefits of PE-ALD HZO films diminish when deposited on a TiN BE, whereas Th-ALD HZO films exhibit reasonably good performance at high operating temperatures when integrated with TiN BE, offering a reliable option for TiN-based high-temperature devices. Furthermore, an oxidized W interface provides significantly greater endurance enhancement than an oxidized TiN interface under comparable oxidation conditions. These insights into the interplay between deposition technique, electrode material, and high-temperature reliability provide important guidelines for integrating ferroelectric devices into advanced 3D systems, BEOL processes, and DRAM/HBM applications operating at their qualification temperatures.

\section*{Acknowledgement}
This work was supported by SUPREME, one of the seven SRC-DARPA JUMP 2.0 centers. Fab was done at the IMS, supported by the NSF-NNCI program (ECCS- 1542174).
\newpage
\ifCLASSOPTIONcaptionsoff
  \newpage
\fi

\end{document}